\documentclass{article}

\usepackage{amsmath,mathtools,amssymb,amsfonts,amsthm,graphicx,color,verbatim,enumerate,dsfont}
\usepackage{setspace}

\usepackage[numbers]{natbib}

\definecolor{Blue}{rgb}{0.3,0.3,0.9}
\usepackage[margin=1.707in]{geometry}

\definecolor{e-mail}{rgb}{0,.40,.80}
\definecolor{reference}{rgb}{.20,.60,.22}
\definecolor{citation}{rgb}{0,.40,.80}
\usepackage[colorlinks, linkcolor=reference,
            citecolor=citation,
           urlcolor=e-mail]{hyperref}

\numberwithin{equation}{section}

\theoremstyle{remark}

\begin{document}

\title{A Model for Innovation Diffusion with Intergroup Suppression\protect}
\author{Anirban Chakraborti, Syed S. Husain
 \\\small School of Computational and Integrative Sciences\\[-0.8ex]
\small Jawaharlal Nehru University, New Delhi 110067, India\\
\\
Joseph Whitmeyer
 \\\small Department of Sociology\\[-0.8ex]
\small University of North Carolina at Charlotte, Charlotte, NC 28223\\}

\maketitle

\begin{abstract}
We present a new model for the diffusion of innovation.  Here, the population is segmented into distinct groups.  Adoption by a particular group of some cultural product may be inhibited both by large numbers of its own members already having adopted but also, in particular, by members of another group having adopted.  Intergroup migration is also permitted.  We determine the equilibrium points and carry out stability analysis for the model for a two-group population.  We also simulate a discrete time version of the model.  Lastly, we present data on tablet use in eight countries from 2012-2016 and show that the relationship between use in the ``under 25" age group and ``55+" age group conforms to the model.  \end{abstract}

\section{Introduction}

Modeling the diffusion of innovation, which can be technological or cultural, has an extensive history, over decades and across disciplines (see, e.g., Coleman \textit{et al.} (1966); Rogers (2003); Mahajan and Peterson (1985)).  One process absent from all innovation diffusion models is suppression by one group of adoption or use of a cultural item in another group.  Yet, as we describe below, this process clearly occurs for some items and has been recognized in some theoretical and empirical work.  We present, analyze, and preliminarily test here, therefore, such a model, adapted from the Bolker-Pacala model in population biology.

People may adopt or abandon cultural items for a variety of reasons, including intrinsic value to them, identity signals to themselves or others, and social pressure.  There also may be external coercion applied.  It has long been noted in social science that adoption of some cultural item may depend on its use by others.  This effect can be positive for many reasons (Leibenstein, 1950; Lieberson, 2000) but also negative, in, for example, what Leibenstein called the snob effect (Leibenstein, 1950).  More recently, and more specifically relevant to the process we are newly including here, Berger and colleagues (Berger \& Heath, 2008; Berger \& Le Mens, 2009) have pointed out that adoption of a cultural item by one group may induce members of another group to abandon it.  Berger and Heath (2007, 2008) make a strong case for this for many cultural items, such as clothing brands and kind of automobile, explaining this effect with an identity motivation (Berger \& Heath, 2007). In \emph{Freakonomics}, Levitt and Dubner (2005) allege this phenomenon for first names, claiming that their California names data show that lower classes adopt names that the higher classes are using, but then that the higher classes abandon those names because the lower classes are now using them. 

Note that in the above examples the negative effect is an internal phenomenon, in that the suppressing group is not trying to lower use and adoption by the other group.  Historically, however, external suppression also has occurred.  One clear example is sumptuary laws, for example, in the Middle Ages in Europe, wherein clothes of certain colors and materials were not permitted to people below a certain social status.  Some barriers to entry, such as requiring men who join the cavalry to come with a horse or charging high fees at golf courses can be seen also a as a higher socioeconomic class keeping certain cultural items at a low level in a lower socioeconomic class.  In the case of external suppression, the reason for lower adoption and use will be different from when the phenomenon is internal; it will not be due to identity motivation, for example.  For our purposes, however, constructing a model incorporating this negative effect, the mechanism is not important.

As yet, models of the diffusion of innovation have not included such negative effects of use by one group on adoption and use by another group.  Early models naturally were simplest, assuming a single homogenous population with a single mode of diffusion.  These were made more complex, again, in various natural ways.  Populations were made heterogeneous, sometimes by positing segmented populations, such that the process works differently in different segments, sometimes modeling a continuous distribution of the population in characteristics or adoption propensity.  Different sources of diffusion were considered, for example, other people or media, and in some models, such as the Bass model (Bass, 1969), different sources were combined.  Bartholomew added a loss-of-interest mechanism to those in the Bass model and also presented a stochastic model, in contrast to the common deterministic models (Bartholomew, 1976).  Alternative assumptions were made concerning the underlying mechanism of diffusion, such as social contagion simply through exposure, social pressure or conformity, and social learning (Young 2009).  Different channels of diffusion also have been examined, for example, with new attention to online diffusion (Go\"{e}l \textit{et al.}, 2012).

One impetus behind the proliferation of models was that not all data showed the same pattern.  For example, the archetypical innovation diffusion pattern is the logistic curve, with an adoption rate peak somewhere in the middle of the diffusion process.  But for some products, adoption was bimodal, with an early rate of adoption peak, then a lull with relatively few new adoptions, then another rate of adoption peak.  This required new models (Karmeshu \& Goswami, 2001, 2008; Karmeshu \& Sharma, 2004).  To state a general principle, different cultural products may differ in their underlying diffusion mechanisms, and so may have fundamentally different diffusion patterns and require different models.

Our new model is presented in the spirit of this principle.  Similar to previous models, it relies on something like contagion as part of the underlying mechanism.  It also is a mean field model, i.e., works with variables aggregated above the individual level, and it divides the population into different segments.  The crucial novelty of this model is that adoption of the product by one segment may have a negative effect on adoption by another segment.  Adoption by one segment is also allowed to have a positive effect on adoption by another segment and excessive adoption within a segment may have a negative effect on further adoption within the segment.  These three effects we call ``external suppression," ``external stimulation," and ``internal suppression," respectively.  In the model, they are all mathematically second-order effects.

We present as an example and test this model on tablet use from 2012-2016.  In numerous countries, tablet use increases in the oldest age group (55+) and at first increases even faster in the youngest age group (under 25).  It appears, however, that when the oldest age group use reaches a certain level, somewhere close to 30 percent, tablet use in the youngest age group starts to decrease.  A plausible mechanism is that when use among older adults gets sufficiently high, the young begin to perceive tablets as an older person’s device, or at least not something that can differentiate them from older people.  Consequently young people become less likely to adopt and some even stop using the device.  This may well be motivated by identity considerations, consistent with Berger and Heath's (2007) argument, but, again, it is not necessary to know the exact mechanism to model its population effect.

For this process, we borrow an appropriate, existing model that has been adapted from the Bolker-Pacala model of population dynamics (Bolker \& Pacala, 1999; Bolker \textit{et al.}, 2003).  We describe the model in the second section of the paper, summarizing the relevant theoretical finding.  In the third section we present the empirical data on tablet use in the form of plots, along with plots of simulations that produce qualitatively similar patterns.  The fourth section summarizes and draws conclusions.


\section{The Model}
We introduce the following model of innovation diffusion, which we call the BP model of innovation diffusion because it is taken from a multi-group mean field approximation of the Bolker-Pacala model of population dynamics in biology (Bessonov \textit{et al.}, 2014, 2016).  The population dynamics model posits an initial population of individuals living on a lattice, i.e., a multi-dimensional grid.  The lattice can represent geographical space, its typical biological use, but it also can represent other spaces on which a population may be distributed.  For example, it could be a one-dimensional space of age or a multidimensional space with dimensions of age, ethnicity, various socioeconomic status measures, and so forth (see, e.g., McPherson, 1983).  Each individual can give birth to a another individual or die or migrate, all at certain rates.  In addition to their intrinsic rates, the existence of individuals may be affected negatively (suppressed) by the presence of other individuals.   A mean field treatment of this is mathematically tractable, and, in fact, is equivalent to a kind of random walk.  In the multi-group version, the population is partitioned into $N$ different groups.  Suppression can occur both within a group and across groups.  

Before exposition of the BP model, let us discuss why the mathematical models we present are useful, that is, why we might want to develop and analyze them instead of, for example, simply looking at a simulation of individuals making certain probabiistic choices.  We present two versions of the BP model, one a stochastic version, equivalent to a random walk, and the other a system of differential equations giving a deterministic trajectory, together with other equations describing the fluctuations around that trajectory.  We can begin by noting that the random walk is exactly equivalent to the simulation of individuals making probabilitistic choices.  Nevertheless, by casting it as a random walk we gain the ability to use the theoretical apparatus that has been developed for random walks, such as the conditions under which it approaches a steady state distribution and other outcomes that we do not develop here.  Analyzing it as stochastic fluctuations around a deterministic solutions to a system of differential equations allows us to identify and classify equilibria, to precisely partition the parameter space with regard to equilibria, that is, the likely fate of the process, and even to note the possibility of interesting rare events such as a large fluctuation pushing the system from one equilibrium to another.

To model innovation diffusion, the initial population consists of the initial adopters of the cultural product.  Adoption of the product by a new person corresponds to birth and abandonment of the product correponds to death.  Suppression within and across groups can inhibit further adoptions or even reduce use of the product within a group.  Migration corresponds to movement by an adopter from one group to another.  

The continuous time model may be presented as follows (Bessonov \textit{et al.}, 2016).  Represent the number in each group $Q_{i,L}$, $i=1,\ldots,N$, at time $t$ who have adopted the cultural product by 
\begin{align}\label{RW}
\boldsymbol{n}(t) = \{n_1(t),n_2(t),\ldots,n_N(t)\},
\end{align}a continuous time random 
walk on $(\mathbb{Z}_{+})^{N}$ with rates obtained from, for $i,j = 1,2,\ldots,N$
\begin{align}\label{DE1}
 \boldsymbol{n}(&t + dt | \boldsymbol{n}(t)) \\ \notag&= \boldsymbol{n}(t) + 
\begin{dcases}
 \, e_i & \text{w. pr. }  \beta_i n_i(t)dt + o(dt^2)\\
-e_i & \text{w. pr. } \mu_i n_i(t)dt  +\frac{n_i(t)}{L}\sum_{j=1}^{N}a_{ij}n_j(t)dt + o(dt^2)\\
e_{j} - e_{i} & \text{w. pr. } n_i(t)q_{ij}dt + o(dt^2), \quad j\neq i \\  
 0 & \text{w. pr. } 1- \sum_{i=1}^{N}(\beta_i+\mu_i)n_i(t)dt \\
 & \qquad \qquad - \frac{1}{L}\sum_{i,j}n_i(t)n_j(t)a_{ij}dt +\sum_{i,j}n_i(t)q_{ij} + o(dt^2)\\
 \text{other } & \text{w. pr. } o(dt^2)
\end{dcases}
\end{align}
where $e_i$ is the vector with $1$ in the $i^{th}$ position and $0$ everywhere else.  

Let us define the variables and parameters.  $\beta_i$ is the adoption rate and $\mu_i$ is the abandonment rate.  The subscript means that they may vary by group.  The multiplication of $\beta_j$ by $n_i$ fits the mechanism being contagion or exposure: it depends on the number who have adopted already.  The multiplication of $\mu_j$ by $n_i$ is because it is precisely those who have adopted a product who can abandon it subsequently.  $q_{ij}$ is the rate of migration from group $i$ to group $j$.  Whether this is possible depends on the nature of the groups.  For example, if they are adjacent age groups, then a positive migration rate from the younger to the older group is inevitable, but the reverse rate must be 0.  In contrast, if the groups are social classes, then movement between all social classes, which is likely, would be conveyed by all migration rates being positive.  The parameter $a_{ij}$ is the rate of supression of group $j$ by group $i$, where $i$ and $j$ can be the same.  Finally, $L$ is adoption capacity, a control for the total number that can adopt, in other words a scale parameter.  Table 1 lists the model parameters together with their meanings. 

\begin{table}[ht]
\centering
\caption{Model Variables and Parameters.}
\begin{tabular}{cc}
\hline
\hline
Parameter & Meaning \\
\hline
$n_i$ & Number of adoptees in group $i$ \\
$\beta_i$ & Adoption rate in group $i$ \\
$\mu_i$ & Abandonment rate in group $i$  \\
$a_{ii}$ & Competition or self-limiting rate for group $i$ \\
$a_{ij}$ & Rate of suppression of group $j$ by group $i$ \\
$q_{ij}$ & Migration rate from group $i$ to group $j$ \\
$L$ & Adoption capacity \\
\hline
\end{tabular}
\end{table}

The equation \ref{DE1} allows us to construct a system of differential equations, but it is convenient to normalize the number of adoptees by dividing by $L$.  We set $$z_i(t) := \frac{n_i(t)}{L}, \qquad i=1,\ldots,N.$$
and define, for $i=1,\ldots, N$
	\begin{align}\label{defF}
   F_i(\mathbf{z}(t)) = 
   \left(\beta_i - \mu_i - \sum_{j\neq i}q_{ij}\right)z_i - a_{ii} z_i^2 - 
   \displaystyle\sum_{j\ne i}a_{ji}z_iz_j + \sum_{j\neq i}q_{ji} z_j. 
	\end{align}
Then, the normalized system of differential equations is
	\begin{equation}\label{dfs2}
	\frac{d \mathbf{z}(t)}{dt} = \mathbf{F}(\mathbf{z}(t)).
	\end{equation}
An equilibrium for the system occurs precisely at the points where
\begin{equation}\label{dfs3}
	\mathbf{0} = \mathbf{F}(\mathbf{z}),
	\end{equation}
with one solution being $\mathbf{z} \equiv \mathbf{0}$.  This process has a functional Law of Large Numbers and functional Central Limit Theorem, that is, as $L \to \infty$ the process converges to a Gaussian diffusion (Bessonov \textit{et al.}, 2016; Kurtz, 1971).  What this means is that for reasonably large $L$ the  process will be very close to the following.  There is a central tendency that is a deterministic trajectory, given by the system of partial differential equations \ref{dfs2} together with an initial value $\mathbf{z_0}$ (see eq. \ref{DE4} for the system for our two-group model).  Because of the stochasticity of the process, however, the values at any given time $s$ will be normally distributed about that deterministic trajectory, with $N \times N$ covariance matrix $\mathbf{G}(\mathbf{z}(t))$ (Bessonov \textit{et al.}, 2016; Kurtz, 1971)
\begin{equation}\label{cov1}
\mathbf{G}(\mathbf{z}(t)) = 
\begin{dcases}
 \,  G_{ii}(\mathbf{z}(t)) = (\beta_i + \mu_i + \sum_{j \ne i} q_{ji}) z_i + \sum_{j} a_{ji} z_iz_j + \sum_{j \ne i} q_{ji} z_j  \\
 \, G_{ij}(\mathbf{z}(t)) = -q_{ij}z_i - q_{ji}z_j  & i \ne j
\end{dcases}
\end{equation}
 This means that from time $s$ to $s+\delta$, for small time increment $\delta$, the covariance matrix will be $G(s)$ multiplied by $\delta$.  For our two-group model, the diffusion is in two dimensions with a $2 \times 2$ covariance matrix.

The normalized system (Eqs. \ref{defF}, \ref{dfs2}, \ref{dfs3}) shows that the process possesses equilibria or steady states, that is, points where the deterministic trajectory remains constant (Bessonov \textit{et al.}, 2016).  From the normalized system, we can find the equilibria and whether the equilibria are stable or unstable, that is, whether when close to an equilibrium the process will approach the equilibrium or not.

In addition, for purposes of simulation, we need transition probabilities for discrete time, which are easily available from equation \ref{DE1}.  Specifically, for $N$ groups, we can simulate the embedded discrete time random walk on $(\mathbb{Z_+})^N$, denoted $\{X_n\}_{n=0}^{\infty}$, associated with the continuous random walk \eqref{RW}. For $\mathbf{x} = (x_1,\ldots,x_N)\in\mathbb{Z_+}^N,$ set 
$$c(\mathbf{x}) = \sum_{i=1}^N \left(\beta_i + \mu_i + \frac{a_{ii}}{L}x_i\right)x_i + \sum_{i,j=1, i\ne j}^N q_{ij} x_i.$$
$\{X_n\}$ has transition probabilities, 
for $\mathbf{x}, \mathbf{y} \in (\mathbb{Z_+})^N$, $\mathbf{x}\neq\mathbf{0}$
\begin{equation}\label{Transition}
P(\mathbf{x}, \mathbf{y}) = \frac{1}{c(\mathbf{x})}\cdot
\begin{dcases}
 \,  \beta_i x_i & \text{if } \mathbf{y} = \mathbf{x} + e_i, i=1,\ldots,N \\
 \, \mu_i x_i + \frac{a_{ii}}{L} x_i^2 & \text{if } \mathbf{y} = \mathbf{x} - e_i, i=1,\ldots,N\\
 \,  q_{ij} x_i & \text{if } \mathbf{y} = \mathbf{x} - e_i + e_j, i \ne j \\
 \, 0 & \text{ otherwise }
\end{dcases}
\end{equation}
Recall that we use $e_i\in\mathbb{Z}^N$ to denote the vector with $1$ in the $i^{th}$ position and $0$ 
everywhere else.

Bessonov \textit{et al.} (2016) shows that a random walk with these transition probabilities is geometrically ergodic.  That is, it is positive recurrent with exponential convergence to a stable distribution.


\subsection{Model for 2 groups}
We now focus on a model restricted to two groups.  Fig. 1 shows a diagram of the random walk for this process starting at a point where $n_1$ members of group 1 have adopted and $n_2$ members of group 2 have adopted.
\begin{figure}[h]
\begin{center}
\scalebox{0.7}{\includegraphics{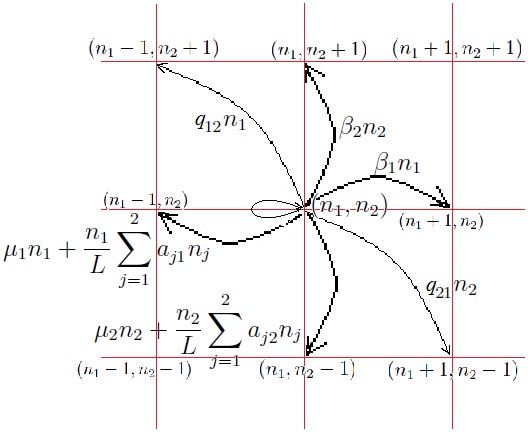}}
\end{center}
\caption{Innovation Diffusion for Two Groups as Random Walk}
\end{figure}

To match our empirical case and keep the calculations straightforward, we assume only one-way migration.  That is, we assume $q_{12}=0$ but allow $q_{21} \geq 0$.  We obtain the equilibria using eqs. \ref{defF}, \ref{dfs2}, \ref{dfs3} and multiplying by $L$ to scale up to population values, or alternatively, by using the following
\begin{equation}\label{DE4}
\begin{dcases}
& \frac{dn_1}{dt} = v_1 n(1) - \frac{a_{11}}{L}n_1^2(t) - \frac{a_{21}}{L}n_1(t)n_2(t) + q_{21}n_2(t)\\
& \frac{dn_2}{dt} = v_2 n(2) - \frac{a_{22}}{L}n_2^2(t) - \frac{a_{12}}{L}n_1(t)n_2(t) - q_{21}n_2(t), 
\end{dcases}
\end{equation}
where, to simplify notation, we use a ``net rate of adoption" for each group by setting $v_i := \beta_i - \mu_i$, for all $i$.

Mathematically, up to four equilibrium points exist, which we label $E_1$, $E_2$, $E_3$, and $E_4$, but in fact four valid points are never realized.  For some configurations of the parameters, only two real singularities exist; the other two are complex and therefore can be disregarded.  For the remaining configuration of parameters, one equilibrium point has a negative value in one of its coordinates.  As that is impossible when the coordinate represents the number of people who have adopted some cultural product, this equilibrium is not valid.

Once the two or three singularities are identified, we carry out a stability analysis by evaluating the Jacobian of the system of differential equations at the different points and using the eigenvalues to classify the kind of singularity in the usual fashion (see, e.g., Logan, 2006).  We will not present an exhaustive description of the kinds of singularities that can exist; that is available in numerous textbooks.  The most important for our purposes are the following.  A stable proper node is a point that a trajectory approaches directly, a focus one that it approaches by spiraling around it.  Unstable versions of these exist: instead of approaching the equilibrium the trajectory leaves it, in the same fashion.  A saddle point is a singularity that a trajectory approaches in one direction but leaves in a different direction (e.g., approaches from the South but recedes heading West); the result is the trajectory makes a more-or-less near pass by the singularity and leaves.

As simple inspection of eqs. \ref{DE4} shows, $E_1 := (0,0)$ is an equilibrium point.  The eigenvalues at $(0,0)$ are $\lambda_1 = v_1$ and $\lambda_2 = v_2 - q_{21}$.  Thus, if the net rate of adoption in group 2 is greater than the rate of migration from group 2 to group 1, that is, $v_2 > q_{21}$, which is by far the most likely scenario, then $(0,0)$ is an unstable proper node; over time, trajectories go away from it.  Should the migration rate be greater, $q_{21} > v_2$, then, $(0,0)$ will be a saddle point---still not a point of attractive stability.

A second singularity always exists at $E_2 := (\frac{v_1L}{a_{11}},0)$.  One eigenvalue is $\lambda_1 = -v_1$ and the second is $\lambda_2 = v_2 - \frac{a_{12}v_1}{a_{11}}- q_{21}$.  Here, if the net adoption rate in group 2, $v_2$, is small enough, then, $\lambda_2 < 0$ and this singularity will be an asymptotically stable proper node.  That is, trajectories will converge to this point; use of the cultural product in group 1 will die out.

Two other possible equilibria exist.  They are complicated, involving the complementary square roots of a quadratic equation.  Setting 
	\[ R := \sqrt{(a_{12}q_{21}-a_{21}q_{21}-a_{22}v_1+a_{21}v_2)^2-4(a_{11}a_{22}-a_{12}a_{21})(q_{21}^2-q_{21}v_2)} \]
$E_3$ and $E_4$ are, respectively, observing the $\mp$ and $\pm$
\begin{equation*}
\begin{array}{l}
	 n_1 =  L\left(\frac{\big(a_{21}q_{21}-a_{12}q_{21}+a_{22}v_1-a_{21}v_2 \mp R\big)}{2(a_{11}a_{22}-a_{12}a_{21})}\right), \\
	 n_2 = \frac{L}{a_{22}} \left(v_2-q_{21} + \frac{a_{12}\big(a_{12}q_{21}-a_{21}q_{21}-a_{22}v_1+a_{21}v_2 \pm R \big)}{2(a_{11}a_{22}-a_{12}a_{21})}\right).
\end{array}
\end{equation*}

\noindent \emph{Simplification} Let us simplify the situation by assuming only one-way suppression, namely that $a_{21} = 0$.  This corresponds to the empirical application in the next section.  In this case, singularities $E_3$ and $E_4$  never can exist together.  Either both are complex or one takes a negative value for one of its coordinates.  In fact, $E_3$ can never be positive in both of its coordinates, so $E_3$ is not a viable equilibrium point.  

$E_4$ can be a viable equilibrium point, however.  If so, it can be either a stable spriral, or an asymptotically stable proper node.  Either way, $E_2$ has to be a saddle point and $E_1$ has to be an unstable proper node.   If neither $E_3$ nor $E_4$ are viable singularities, then $E_1$ can be either an asymptotically stable proper node or a saddle point and $E_2$ is an asymptotically stable proper node.  Table 1 summarizes the three possible configurations of singularities, along with a simple necessary condition.  The full conditions distinguishing the first and second singularity are complicated and so omitted from the table.

\[
\begin{array}{ccccc} \mathrm{Necessary} & E_1 & E_2 & E_3 &  E_4 \\
\mathrm{Condition} \\
\hline 
v_2 > q_{21} & \text{unstable} & \text{saddle point} & \text{not viable} & \text{asymptotically} \\
 & \text{proper node} & & & \text{stable focus} \\
\\
v_2 > q_{21} & \text{unstable} & \text{saddle point} & \text{not viable} & \text{asymptotically} \\
 & \text{proper node} & & & \text{stable proper node} \\
\\
v_2 \leq q_{21} & \text{saddle point} & \text{asymptotically} & \text{not viable} & \text{not viable} \\
 & & \text{stable proper node} \\
\hline
\end{array}
\]
\noindent\hspace{1 cm} Table 1. Possible Equilibrium States for Innovation Diffusion Model.


\subsection{Two group example with stable positive equilibrium}
As an example of the innovation diffusion process, and to provide a comparison with the empirical data from the next section, we present the model outcomes for parameter settings chosen in the range in which there is a stable positive equiibrium.  Specifically, we use as parameters the following values: initial values, $n_1(0) = n_2(0) = 10$; scale, $L=1000$; adoption rate and drop rate for group 1, $\beta_1 = .0003$, $\mu_1 = .0001$, for group 2, $\beta_2 = .0006$, $\mu_2 = .0001$; internal suppression, $a_{11}=.0002$, $a_{22} = .0001$; suppression (inhibition) of group 2 by group 1, $a_{12} = .0003$; migration from group 2 to group 1, $q_{21} = .00005$.

This approaches an Ornstein-Uhlenbeck process.  In other words, there will be stochastic fluctuations about the mean trajectory given by the system of differential equations in eqn. \ref{DE4}, and once the trajectory nears the equilibrium point these fluctuations will be distributed normally.  There, the trajectory will have local drift $\mathbf{F}'(E_4)$ (see eq. \ref{defF}) and local covariance matrix $\mathbf{G}(E4)$ (see eq. \ref{cov1}).  
We also carried out a discrete time simulation of the innovation diffusion process with the saame parameters.  This used the embedded random walk with transition probabilities given in eq. \ref{Transition}.  Fig. 2 shows the trajectories of the numbers of adoptees in the first and second groups.
\begin{figure}[h]
\begin{center}
\scalebox{0.9}{\includegraphics{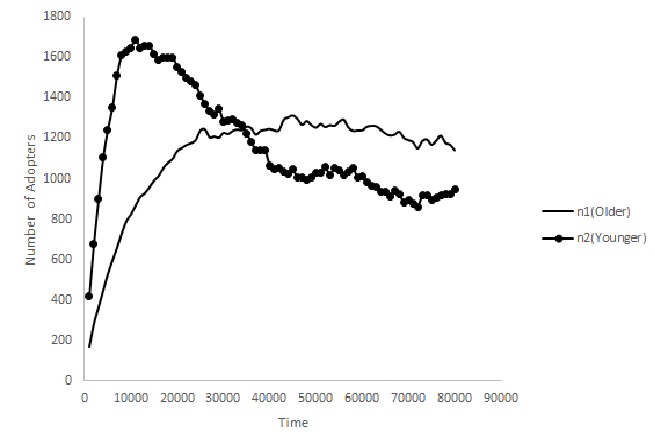}}
\end{center}
\caption{Trajectories of Simulation of Innovation Diffusion for Two Groups}
\end{figure}
Another way of looking at the outcome is the projection of the trajectory in the $n_1 - n_2$ plane, showing how the numbers of adoptees in each group relate to each other.  Labeling the groups ``older" and ``younger" to match the empirical examples to follow, one such simulated trajectory (the wobbly curve) is shown in Fig. 3, together with the numerical solution of the differential equation system, eq. \ref{DE4}, with the same parameters and initial conditions (the smooth curve).  In the simulated trajectory, the number of adoptees in the older group ($n_1$) increases monotonically in time, so that time may be taken as increasing from left to right.  The presentation of the differential equation system solution is parametric, so that time increases as the smooth curve proceeds away from the origin.  Clearly, the simulation produces a stochastic path close to the smooth deterministic path given by the differential equation system.
\begin{figure}[h]
\begin{center}
\scalebox{0.4}{\includegraphics{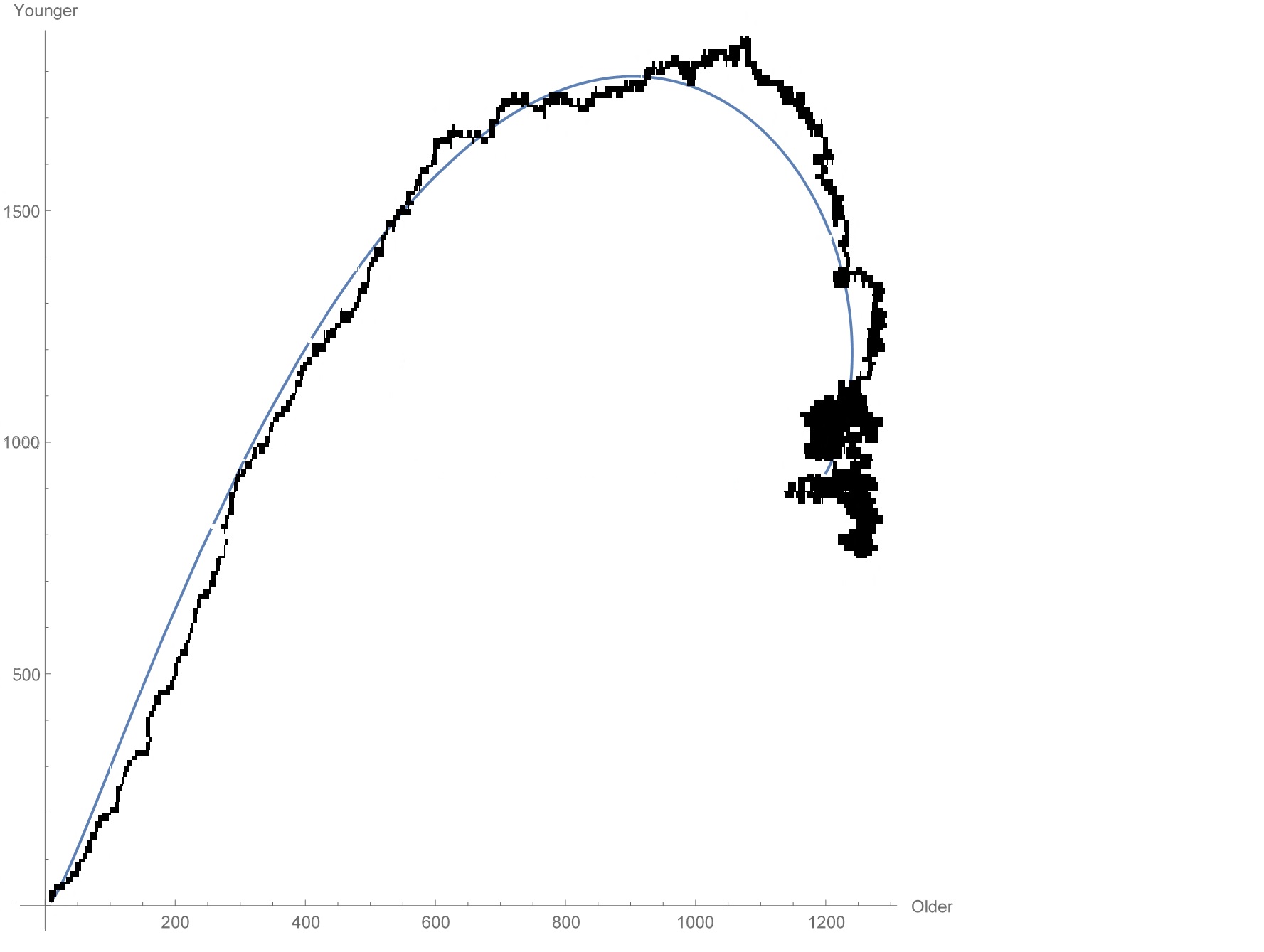}}
\end{center}
\caption{Simulation of Trajectory for Innovation Diffusion for Two Groups}
\end{figure}
The stability analysis for the model with these parameters gives three singularities, an unstable proper node at $E_1 = (0,0)$, a saddle point at $E_2 = (1000,0)$, and an in-spiral at $E_4 = (1193, 921)$; this last corresponds to the limit point of the deterministic curve in Fig. 3.  The classification of $E_4$ follows because the Jacobian has one positive eigenvalue and one negative eigenvalue at $E_4$.  The simulated trajectory clearly conforms to the theoretical analysis.  At the equilibrium point, the Gaussian diffusion has local diffusion $\mathbf{F}'(E_4) = (-.48, -.54)$.  That the drift is negative means that the farther it deviates from the equilibrium point the more stronger the trajectory will be pulled toward the equlibrium point.  The local covariance matrix is 
\[\mathbf{G}(E_4)=\left[\begin{array}{cc}
761.9 & -.046 \\
-.046 & 1059.2
\end{array}\right]\]


\section{Empirical Data on Tablet Use}
We present data for a situation that we suggest corresponds to the scenario being modeled.  The cultural product in question is the tablet (computer), and the groups in question are age groups.  We focus on the youngest age group, ``under 25," and the oldest age group, ``55+."  We suggest that the youngest group will have a greater net adoption rate than the older age group, due to characteristics such as greater long-term expected payoffs to adopting new technology and greater intensity of social contacts, which facilitates the spread of information and influence.  We also assume, however, that if use of tablets in the older group rises too high, the younger group will begin to perceive the device as something for older people, at least not special for younger people; the tablet will lose much of its status value for younger people and this will inhibit or suppress its use.  We assume there is no corresponding suppression of use by older people due to use by younger people.  Finally, while there clearly is no direct migration from the ``under 25" group to the ``55+" group, there will be migration from ``under 25" to ``25 - 34," from ``25 - 34" to ``35 - 44," from ``35 - 44" to ``45 - 54," and from ``45 - 54" to ``55+."   This we make take to be indirect migration from the youngest to the oldest age groups.

Figures 4 through 9 show tablet use from 2012 through the first half of 2016 in 6 countries, with data from the Google Consumer Barometer (2016).  These graphs the youngest group use against the oldest group use; the third dimension, time, is omitted.  It may be noted, however, that tablet use in the oldest age group increases monotonically with time in the oldest age group.  Thus, in each graph, time increases from left to right.

\begin{figure}[h]
\begin{center}
\scalebox{0.7}{\includegraphics{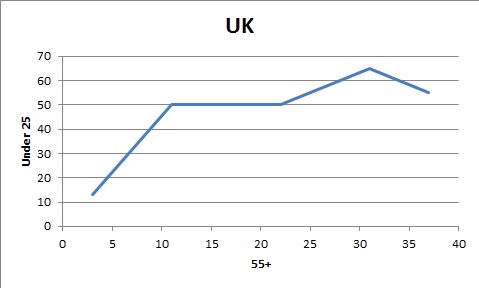}}
\end{center}
\caption{Tablet Use in UK for Under 25 and 55+ Age Groups, 2012 - 2016.}
\end{figure}
\bigskip

\newpage
\begin{figure}[h]
\begin{center}
\scalebox{0.7}{\includegraphics{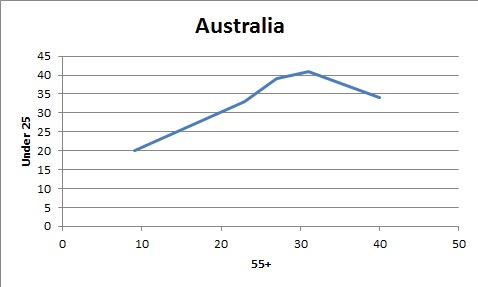}}
\end{center}
\caption{Tablet Use in Australia for Under 25 and 55+ Age Groups, 2012 - 2016.}
\end{figure}
\bigskip

\begin{figure}[h]
\begin{center}
\scalebox{0.7}{\includegraphics{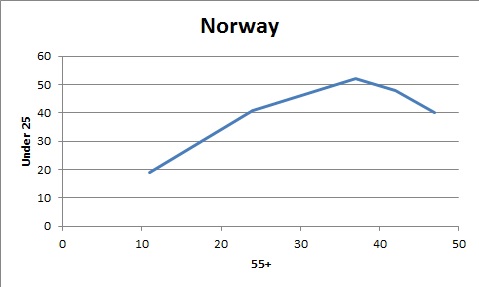}}
\end{center}
\caption{Tablet Use in Norway for Under 25 and 55+ Age Groups, 2012 - 2016.}
\end{figure}
\bigskip

\newpage
\begin{figure}[h]
\begin{center}
\scalebox{0.7}{\includegraphics{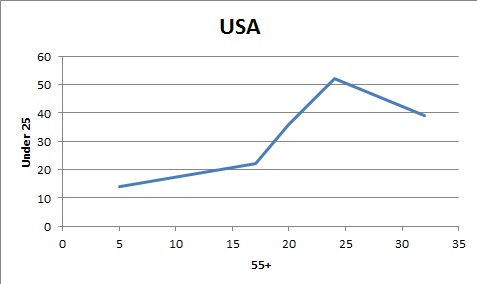}}
\end{center}
\caption{Tablet Use in US for Under 25 and 55+ Age Groups, 2012 - 2016.}
\end{figure}
\bigskip

\begin{figure}[h]
\begin{center}
\scalebox{0.7}{\includegraphics{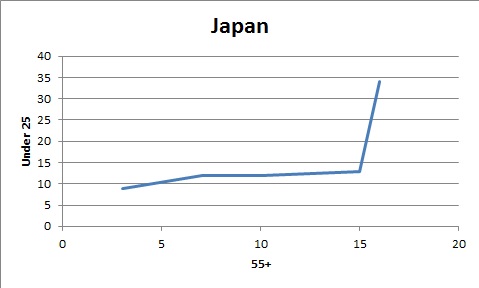}}
\end{center}
\caption{Tablet Use in Japan for Under 25 and 55+ Age Groups, 2012 - 2016.}
\end{figure}
\bigskip

\newpage
\begin{figure}[h]
\begin{center}
\scalebox{0.7}{\includegraphics{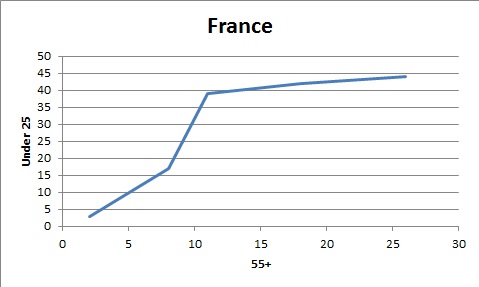}}
\end{center}
\caption{Tablet Use in France for Under 25 and 55+ Age Groups, 2012 - 2016.}
\end{figure}
\bigskip

All six countries show that the increase in tablet use initially progresses more quickly in the younger age group.  In four countries, clearly, tablet use ultimately declines in the younger group along with the last rise in the oldest group.  This is not the case in two countries, Japan and France.  The plots of the deterministic trajectory and of the simulation shown in Fig. 3 resemble the empirical graphs of Figs. 4-9.  Recall that for those plots, the parameters speciify a higher net adoption rate in the younger group, a very small migration rate from the younger to the older group, and external suppression from the older group to the younger group but not the other direction.  Concerning Japan and France, note that in these two countries tablet use in the oldest age group has not reached the levels that it has in the other four countries.  Thus, arguably the model may apply to these two countries as well, they are just at an earlier portion of the innovation diffusion process.

	The results of a goodness-of-fit test support this interpretation as well as the applicability of the model more generally.  We tested the goodness-of-fit of the empirical data for tablet use to predictions of the simulation model, using the same parameter settings as given above.  In fact, the model predictions used were those from the simulation run depicted in Figs. 2 and 3.  Table 2 below shows the results using a chi-squared goodness-of-fit statistic.  For each country, the maximum values are equated, which calibrates the data, then the simulation gap corresponding to one year (usually 4,000 iterations) is estimated, and finally a starting time in the simulation is estimated.  This leaves seven degrees of freedom. 

	Clearly, for all six countries the model fits somewhat; we cannot reject the null hypothesis that the model fits the data ($p > .05$).  It may be that a simpler model, namely a linear one, fits two of the countries, France and Japan, just as well.  Nevertheless, the BP model fits sufficiently well the data from all six countries. 

\begin{table}[ht]
\centering

\caption{Fit of Empirical Data on Tablet Use, 2012 - 2016, to Simulation Model Outcomes.  Chi-squared Statistic with Seven Degrees of Freedom.}
\begin{tabular}{cc}
\hline
\hline
Country & Chi-squared \\
\hline
United Kingdom & 12.51$^*$ \\
United States & 9.31$^{**}$ \\
Australia & 13.51$^*$ \\
Norway & 9.78$^{**}$ \\
France & 12.17$^*$ \\
Japan & 13.81$^*$ \\
\hline
$^*p>.05,$ $^{**} p>.1$
\end{tabular}
\end{table}


\section{Conclusion}
We have presented here a new model of the diffusion of innovation.  This is a model for a population divided into different groups, where adoption and use of the cultural product by one group may be negatively affected by use by a different group.  The mathematics of the model is taken from the multi-layer Bolker-Pacala model of population dynamics.  We present empirical evidence from several countries for tablet use that conforms to a pattern generated by the model, as shown by a simulation and supported by goodness-of-fit tests.

The empirical pattern of tablet use, with quck adoption in one group but then decline, while the adoption in another group is slower, without decline, is unusual.  Many models of the diffusion of innovation have been developed but none have been applied to such a situation, hence, the need for at least one more model.  We do not claim that this BP model is generally appropriate, but we suggest that in situations of external suppression and inhibition, that is, from one group vis-a-vis another, this model can work well.  Our empirical analysis here was for tablet use but we noted above other examples of this phenomenon in the literature such as first names and automobile makes, as well as historical examples such as sumptuary laws.

We might note that the BP model can quickly present analytical difficulties. With age groups, fortunately, migration can occur in only one direction, but with other sorts of segmentation of populations, say along social class or region, migration would be possible in both directions.  Even this small complication makes analyzing the steady states much more difficult.  Considering more than two groups also would be desirable but, again, this greatly raises the level of analytical difficulty.  It is always possible to simulate more complicated models, but a mathematical analysis is valuable for providing understanding.  For example, in section 3, through finding the singularities and evaluating the Jacobeans at the singularities, we gain a fairly throrough understanding of the dynamic system, what its tendencies are, and how these are affected by the parameters.



\end{document}